\documentclass[10pt,conference]{IEEEtran}
\usepackage{cite}

\usepackage{amssymb}
\usepackage{subcaption}
\usepackage{xcolor}

\ifCLASSINFOpdf
  \usepackage[pdftex]{graphicx}
  \graphicspath{{../pdf/}{../jpeg/}}
  \DeclareGraphicsExtensions{.pdf,.jpeg,.png}
\else
   \usepackage[dvips]{graphicx}
   \graphicspath{{../eps/}}
   \DeclareGraphicsExtensions{.eps}
\fi
\usepackage{amsmath}
\usepackage{algorithmic}

\usepackage{array}

\usepackage{comment}

\usepackage{listings}
\usepackage{graphicx}

\newcommand{\redHL}[1]{{\textcolor{red}{#1}}}

\usepackage[font=small]{caption}

\makeatletter
\newcommand{\linebreakand}{%
  \end{@IEEEauthorhalign}
  \hfill\mbox{}\par
  \mbox{}\hfill\begin{@IEEEauthorhalign}
}
\makeatother

\begin{document}
%
\title{Spec2Cov: An Agentic Framework for Code Coverage Closure of Digital Hardware Designs}

\author{\IEEEauthorblockN{Sean Lowe, Elias Hilaneh, Alma Babbit, Nakul Gopalan, Vidya Chhabria, Aman Arora}
\IEEEauthorblockA{Arizona State University, Tempe, Arizona, USA\\
Email: slowe8@asu.edu} \vspace{-10mm}
}


%


\bstctlcite{BSTcontrol}

\maketitle

\begin{abstract}
Hardware verification is one of the most challenging stages of the hardware design process, requiring significant time and resources to ensure a design is fully validated and production-ready. Verification teams aim to maximize design coverage while ensuring correct behavior and alignment with the specification. Coverage closure, which relies on iterative constrained-random and directed testing, is still largely manual and therefore slow and labor-intensive.
Recent advances show that the code generation capabilities of Large Language Models (LLMs) can be integrated with external tools to build agentic workflows that autonomously perform hardware design and verification tasks. In this work, we introduce Spec2Cov, an agentic framework that automatically and iteratively generates test stimulus directly from design specifications to accelerate coverage closure. Spec2Cov coordinates interactions between an LLM and a hardware simulator, managing compilation and simulation errors, parsing coverage reports, and feeding results back to the model for refinement.
We  present features that improve Spec2Cov’s effectiveness without additional fine-tuning and evaluate their impact. Across 26 designs of varying size and complexity, including problems from the CVDP benchmark suite, Spec2Cov achieves 100\% coverage on simpler designs and up to 49\% on more complex designs.
\end{abstract}

\section{Introduction} \label{sec:intro}
Pre-silicon hardware verification is an exceedingly laborious and time-intensive process, often representing a critical bottleneck in achieving tapeouts. 
The goal of verification is to ensure that the implemented design matches the specifications.
While faster execution is crucial, the necessity for high-quality verification, characterized by high coverage and low test fail rates, cannot be overstated. 


Coverage closure is an iterative process that starts by interpreting design specifications, drafting a detailed testplan, and developing a testbench with constrained-random testcases.
Coverage data from regression runs is analyzed and directed tests are written to iteratively close coverage holes. In current practice, this  cycle is still largely manual and consumes a substantial portion of verification time.

\begin{figure}[!htbp]
    \centering
    \includegraphics[width=\linewidth]{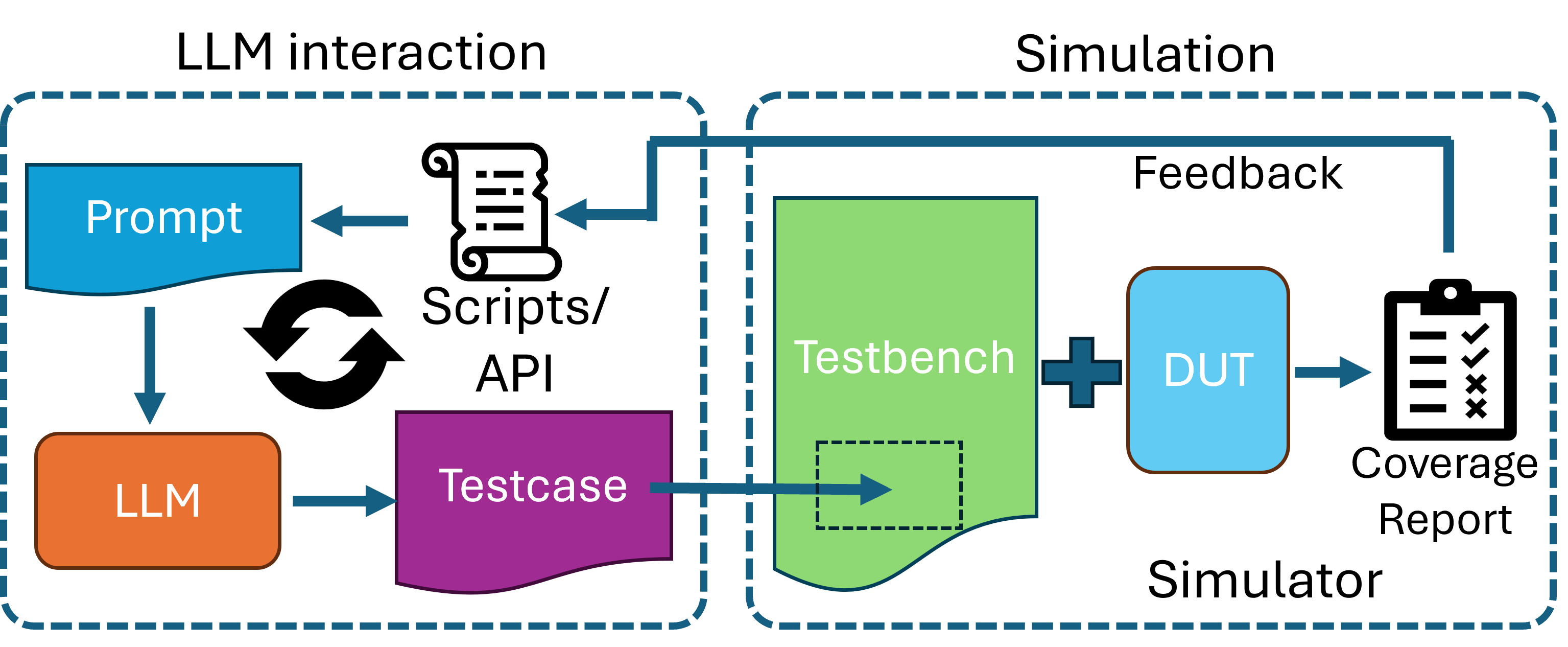}
    \caption{An overview of Spec2Cov, our agentic framework for rapid coverage closure.}
    \label{fig:framework-flow}
\vspace{-3mm}
    
\end{figure}

Recent research has demonstrated that large language models (LLMs) can be employed to generate design and verification code automatically \cite{blocklove:chipchat:2023,ma:verilogreader:2024a,qiu:autobench:2024a,zhang:llm4dv:2023}, including assertions and testcases. 
However, prior work has largely overlooked using design specifications as the primary source for verification code generation, which reduces practical applicability. Showing design code early can bias the LLM and leave design bugs unchecked.
The design code is best shown to the LLM only during the iterative coverage closure phase.
Moreover, the emergence of agentic frameworks - systems that combine LLMs with external tools and feedback loops - has opened new avenues for automation. 
Leveraging context-awareness, reasoning, and agentic capabilities of modern LLMs, we can significantly reduce human effort for coverage closure.




In this work, we present Spec2Cov, an agentic framework that mirrors this workflow with minimal human intervention. It generates testcases directly from design specifications, integrates them into a testbench template, starts with constrained-random testing, and iteratively refines stimuli using coverage reports and simulator errors to close code-coverage holes.

While functional coverage offers a more comprehensive assessment of design behavior and can improve overall quality, code coverage is widely regarded as essential across the industry and is typically considered the minimum requirement for verification signoff. 
Given its critical role, Spec2Cov framework focuses on code coverage, specifically statement (or line) coverage. Nevertheless, the proposed methodology is general and can be easily extended to other types of code coverage and functional coverage.

A high-level overview of Spec2Cov is shown in Fig \ref{fig:framework-flow}. The LLM is given a specification that contains supporting context about a design's functionality, submodules and implementation details. 
The LLM generates testcases which are integrated into a testbench template, starting with constrained random testcases.
Based on simulator feedback, the LLM adapts and generates new testcases, progressively improving coverage in an autonomous loop.


Spec2Cov accelerates coverage closure by automating stimulus generation and feedback using LLMs and simulator integration, reducing manual effort and speeding up verification. This frees engineers to focus on higher-value tasks like debugging and architectural refinement, boosting productivity without increasing headcount. It requires no additional model training, leveraging off-the-shelf LLMs for easy adoption, and lays the foundation for future agentic verification workflows.


\begin{table*}[t]
    \centering
    \caption{Comparison of LLM-assisted hardware verification works}
    \label{tab:related-works}
    \begin{tabular}{l|c|c|c|c|c}
        \hline
        \textbf{Work} & \textbf{DV Focus} & \textbf{LLMs Used} & \textbf{Agentic?} & \textbf{Goal} & \textbf{Benchmarks / Size} \\
        \hline
        ChiRAAG \cite{mali:chiraag:2024a} & Assertion gen. & ChatGPT (GPT-4) & Yes & Improve SVA quality & 7 OpenTitan mods; 3--16 SVAs \\
        AssertLLM \cite{fang:assertllm:2024a} & Assertion gen. & ChatGPT+RAG & No & Spec $\rightarrow$ SVAs & 1 I2C design; 23 signals \\
        AutoBench \cite{qiu:autobench:2024a} & Hybrid TB gen. & ChatGPT (GPT-4) & Yes & UVM TB automation & VerilogEval-Human; 156 HDLBits \\
        LLM4DV \cite{zhang2025llm4dvusinglargelanguage} & Stimulus gen. & GPT-3.5, LLaMA, Claude & Yes & Func. cov. closure & 3 designs; $\sim$3.9k bins \\
        Verilog Reader \cite{ma:verilogreader:2024a} & Test gen. & GPT-4 / Turbo & Yes & Code coverage closure & 24 custom designs; up to 500+ LOC \\
        Hassan et al. \cite{hassan:promptverifyrepeat:2025} & Iterative DV & GPT-4 (example) & Yes & General DV loop & No formal dataset (conceptual) \\
        Ye et al. \cite{ye2025conceptpracticeautomatedllmaided} & UVM env. gen. & GPT-4 & Yes & UVM automation & 9 designs; $\leq$1.6k LOC \\
        \textbf{Spec2Cov (Ours)} & Stimulus gen. & GPT-4o & Yes & Code coverage closure & CVDP + GitHub; 26 designs \\
        \hline
    \end{tabular}
\end{table*}

We make the following contributions in this work:
\begin{itemize}
    \item \textbf{Spec2Cov Framework:} We introduce \textit{Spec2Cov}, an agentic framework that generates testcases directly from specifications and automates code coverage closure.

    \item \textbf{LLM Enhancement Techniques:} We develop several techniques to improve the LLM’s effectiveness in generating high-quality testcases and accelerating coverage closure. These include: 
         \textit{Testplan creation} from specifications,
         \textit{Batched generation} for parallel exploration,
         \textit{Context pruning} to manage prompt length.

    \item \textbf{Comprehensive Evaluation:} We evaluate Spec2Cov on 26  designs, including verification tasks from the CVDP \cite{pinckney2025comprehensiveverilogdesignproblems} and some designs from GitHub. These designs span a wide range of complexity levels (easy, medium, hard) and code sizes (14 to 4500 lines). 

    \item \textbf{Performance Analysis:} We assess Spec2Cov using pass@k and code coverage metrics, and report its computational cost in terms of LLM generation time and token usage.
\end{itemize}
Spec2Cov is the first open-source\cite{spec2cov_anonymized} framework focused on  code coverage closure directly from specifications using CVDP designs for demonstration.


\section{Related Work} \label{sec:related_work}

Several studies have illustrated the use of LLMs for the generation of hardware design code.
Chip-Chat \cite{blocklove:chipchat:2023} studies the challenges and opportunities in using LLMs for hardware design. The authors use ChatGPT to generate a small 8-bit microprocessor design. In VeriGen \cite{thakur:verigen:2024}, the generation of hardware designs using ChatGPT and some fine-tuned LLMs is studied using a proprietary dataset created from GitHub and textbooks. GPT4AIGChip \cite{fu2025gpt4aigchipnextgenerationaiaccelerator} applies ChatGPT to AI accelerator design and explores the limitations of LLMs when creating accelerator designs. Most recently, VerilogDB \cite{calzada2025verilogdblargesthighestqualitydataset} and CVDP \cite{pinckney2025comprehensiveverilogdesignproblems} present larger and more challenging RTL datasets for benchmarking and fine-tuning uses.
Another interesting direction is debugging using LLMs. 
MEIC \cite{xu:meic:2025} focuses on debugging RTL design code using fine-tuning LLMs in an agentic workflow.
UVLLM \cite{hu2024uvllmautomateduniversalrtl} identifies syntactical and functional errors in RTL designs using an agentic pipeline. 

There are some works using LLMs for hardware verification as well (Table \ref{tab:related-works}). 
ChiRAAG  \cite{mali:chiraag:2024a} uses ChatGPT to interactively improve verification assertion generation.
AssertLLM \cite{fang:assertllm:2024a} showcases using ChatGPT and Retrieval Augmented Generation (RAG) to generate SystemVerilog assertions using design and specification using only one design.
AutoBench \cite{qiu:autobench:2024a} uses Anthropic Claude to generate hybrid testbenches using a long generation pipeline, using a dataset derived from HDLBits.
LLM4DV \cite{zhang2025llm4dvusinglargelanguage} demonstrated LLMs for hardware test stimulus generation from given functional coverpoints using three designs. 
Verilog Reader \cite{ma:verilogreader:2024a} focuses on code coverage centric testcase generation by providing the design's code to the LLM. It adds blocks such as DUT explainer and coverage explainer blocks to assist the LLM. Simple designs such as multiplexer, ALU, FSMs, counters and arbiters are used for evaluation.
Hassan et al. \cite{hassan:promptverifyrepeat:2025} propose a simple framework to iteratively reach verification goals and review LLMs playing a semantic role in verification. 
Finally, Ye et al. \cite{ye2025conceptpracticeautomatedllmaided} propose an agentic UVM-based framework and a new benchmark suite of nine designs to evaluate their framework.

Our work focuses on generating testcases directly from design specifications  with the goal of code coverage closure. 
\textit{A direct comparison with these works is not feasible due to fundamental differences in methodologies:} (1) Spec2Cov uses a practical approach of providing specifications to generate testcases whereas other works provide the design directly to the LLM which can lead to testcases that have common-mode errors as the design code. (2) Spec2Cov is evaluated with a mix of CVDP benchmarks and open-source designs gathered from GitHub. No other frameworks use CVDP benchmarks. Prior works use much simpler designs, which do not challenge the capabilities of modern LLMs.


\section{The Spec2Cov Framework} \label{sec:proposal}

\begin{figure*}[t]
    
    \centering
    \includegraphics[width=\linewidth]{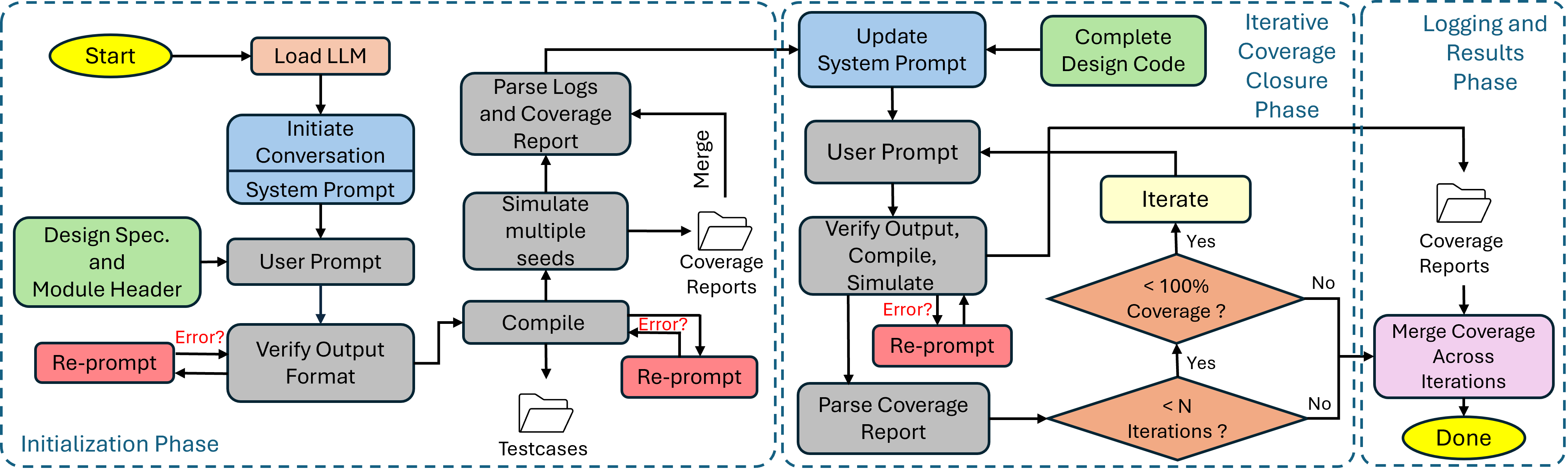}
    \caption{Detailed flow of the Spec2Cov framework}
    \label{fig:spec2cov_details}
    \vspace{-3mm}
\end{figure*}

\subsection{Detailed Operation}

A detailed view of Spec2Cov is shown in Fig. \ref{fig:spec2cov_details}.
Spec2Cov interacts with the LLM through a structured sequence of prompts and with the simulator through a series of commands, while managing communication between them and logging artifacts. The overall process can be split into three phases.

\textbf{Phase 1: Initialization}
The process begins with a \textit{system prompt}, which defines the role of the LLM as a verification engineer and specifies formatting rules for the expected output so that it can be easily parsed and processed. 
This is followed by the first \textit{user prompt} in which the LLM is provided with the design specification and the top module's port declarations extracted from the design. 
The prompt instructs the LLM to use constrained random stimulus, emulating industry-standard verification practices, to maximize initial coverage by generating testcases using constructs such as \texttt{\$urandom} and \texttt{.randomize()}. 
This is referred to as the start of a \textit{conversation}.

If the LLM output cannot be parsed into the expected testcase structure, Spec2Cov marks it as a \textit{decode error}. Upon successful generation, the testcase is inserted into an auto-generated testbench template (which instantiates the design and includes clock generation logic).
The design and testbench are then passed to the simulator, which performs simulation with coverage metrics enabled. Compilation and runtime errors are fed back to the LLM for correction. 
Upon successful simulation of a user-specified number of random seeds, the framework merges the coverage reports. This constitutes the initial achieved coverage.

\textbf{Phase 2: Iterative Coverage Closure.}
Next, the framework parses the coverage report and extracts coverage holes in each module of the design. It then randomly chooses a module with coverage holes for the LLM to target. 
The LLM is shown the design code for the first time in this phase, mirroring the workflow of a human verification engineer. The prompt includes a randomly chosen module's code, with annotations pointing out the coverage holes. 
At this stage, the complete design code is also embedded into the system prompt, enabling the LLM to understand inter-module context. This helps the LLM identify top-level stimuli to effectively target coverage gaps in lower-level modules.

Each interaction in this phase is referred to as an \textit{iteration}. The LLM either generates a new testcase or modifies an existing one to target the identified coverage holes. The updated testbench is simulated, and the new coverage report is fed back to the LLM. This iterative loop continues until either 100\% coverage is achieved or a user-defined maximum number of iterations is reached.
At the end of the process, coverage data from all iterations is merged.

\textbf{Phase 3: Logging and Results}
Spec2Cov logs key artifacts and metrics throughout execution and stores them in a results directory. 
For each conversation, it saves testcases, compilation \& simulation outputs, and coverage databases \& reports. It also produces per-iteration summaries with error counts, coverage metrics, token usage, and runtime.
This supports reproducibility and downstream analysis.

\subsection{Additional Features}
Spec2Cov employs features to improve the performance of the LLM in achieving coverage goals. Each of these features can be enabled or disabled easily through the command line.

\subsubsection{Testplan Creation} \label{subsubsec:directshot}

To more closely mirror industrial verification workflows, we prompt the LLM to generate a detailed testplan (or verification plan). This plan outlines the stimulus and corner cases likely needed to achieve high coverage. By doing so, we provide the LLM with synthetic context that mimics human reasoning, enhancing its ability to generate meaningful and targeted testcases.

\subsubsection{Batched Generation} \label{subsubsec:directshotwtestplan}

Inspired by the way ChatGPT can produce multiple responses for user selection, Spec2Cov supports parallel generation of multiple testcases. The number of testcases is user-defined; we typically use 5, based on empirical observations that the pass@5 probability exceeds 95\% across designs. This means that among five generated testcases, at least one is likely to pass with high coverage. All testcases are simulated, and the one achieving the highest coverage is selected to continue the iterative process.

\subsubsection{Context Pruning}

Maintaining a manageable conversation length is critical when interacting with LLMs. As the token count grows, the model’s ability to utilize context effectively diminishes, and generation time and cost can increase significantly. To address this, Spec2Cov employs \textit{context pruning}, which trims the conversation to stay within a 15,000-token limit. We prune segments where the LLM is correcting its own mistakes, ensuring that only the most relevant and informative context is retained for subsequent iterations.

\section{Methodology} \label{sec:methodology}



{\textbf{LLM Selection:}}
We use ChatGPT-4o as the primary LLM in base form (no fine-tuning or post-training). Spec2Cov is model-agnostic and can run other LLMs with minimal integration effort.
Other models such as Llama 3.3, Mixtral, Mistral, and DeepSeek were evaluated but did not consistently generate syntactically correct Verilog.

{\textbf{LLM Settings:}}
The LLM is configured with specific values of key parameters such as temperature and top-p. 
Temperature controls randomness in generated testcases, and Top-p governs diversity by adjusting nucleus sampling.
We choose a temperature of \textbf{0.3} and a top-p of \textbf{0.7} based on a hyperparameter exploration using pass@5 as an evaluation metric.
These values are selected to balance diversity with syntactic correctness.  

\textbf{Framework and Infrastructure}:
Spec2Cov is able to run LLMs locally or use an API endpoint like OpenAI or Azure, giving users a plug-and-play opportunity. The framework is written in Python and is controllable through various command line arguments that control settings like which LLM to use, which features to enable, how long the conversations are, etc. 
For local generation, the framework uses vLLM \cite{kwon2023efficient} for efficient GPU utilization and batched inference. 

\textbf{Simulation Setup:}
Intel's QuestaSim 23.4 is used to compile, simulate, and extract coverage metrics. 
However, the framework features a simulation API allowing any  simulator to be plugged in. We currently support VCS and Verilator.

\textbf{Design Sources:}
We evaluate our framework on two classes of designs.
(1) \textbf{CVDP\cite{pinckney2025comprehensiveverilogdesignproblems} designs:} 
    These are obtained from the agentic verification related problems in the CVDP benchmark suite (cid12-14). Because the CVDP dataset has many different problems, we use a subset of the suite that has complete design specifications and design code available. 
    There are a couple of designs from the CVDP verification subset that are missing source files, so they were excluded from the subset we used. This resulted in 14 usable designs that have been incorporated.
(2) \textbf{GitHub designs:}
    Since CVDP has only been recently released, we started creating a dataset of open-source designs from GitHub. It was challenging to find designs with specifications and with verification collateral on GitHub. 
    We use twelve designs from GitHub \cite{secworks_vndecorrelator, avashist_fifo, secworks_uart, secworks_sha1, secworks_chacha, secworks_trng, mczerski_sdcard, samidhm_float, utlca_pooling} for evaluating Spec2Cov. We include these designs to increase the complexity diversity of the subset we are experimenting with. Notably, the seven hardest designs we use are all sourced from GitHub. These designs allow us to evaluate the performance of Spec2Cov on harder problems than CVDP.
    

Table \ref{tab:design-difficulty} summarizes the designs by number of levels of module hierarchy, lines of code, and difficulty levels.
We consider a design to be \textit{Hard} for coverage closure if total lines $\geq 450$ or hierarchy $\geq 3$, \textit{Medium} if $150 \leq$ total lines $< 450$ or hierarchy $=2$, otherwise \textit{Easy}.

\begin{table}[t]
\centering
\caption{Designs used for evaluating Spec2Cov}
\small
\begin{tabular}{p{0.3cm} p{3cm} p{1.5cm} p{1cm} p{1cm}}
\textbf{Key} & \textbf{Design} & \textbf{Module Hierarchy Levels} & \textbf{Total Line Count} & \textbf{Relative Difficulty} \\
\hline
e1 & lfsr~\cite{pinckney2025comprehensiveverilogdesignproblems}                        & 2  & 14    & Easy \\
e2 & caesar\_cipher~\cite{pinckney2025comprehensiveverilogdesignproblems}               & 1  & 21    & Easy \\
e3 & dual\_port\_memory~\cite{pinckney2025comprehensiveverilogdesignproblems}           & 1  & 30    & Easy \\
e4 & multiplexer~\cite{pinckney2025comprehensiveverilogdesignproblems}                  & 1  & 35    & Easy \\
e5 & cont\_adder~\cite{pinckney2025comprehensiveverilogdesignproblems}                 & 1  & 51    & Easy \\
e6 & fixed\_arbiter~\cite{pinckney2025comprehensiveverilogdesignproblems}               & 1  & 77    & Easy \\
e7 & alu~\cite{pinckney2025comprehensiveverilogdesignproblems}                          & 1  & 72    & Easy \\
e8 & ttc\_lite~\cite{pinckney2025comprehensiveverilogdesignproblems}                    & 1  & 98   & Easy \\
e9 & vndecorrelator~\cite{secworks_vndecorrelator}  & 1  & 80   & Easy \\
e10 & door\_lock~\cite{pinckney2025comprehensiveverilogdesignproblems}                   & 1  & 137   & Easy \\
e11 & fifo~\cite{avashist_fifo}                     & 1  & 94   & Easy \\
e12 & memory\_scheduler~\cite{pinckney2025comprehensiveverilogdesignproblems}            & 1  & 129   & Easy \\
m1 & sorter~\cite{pinckney2025comprehensiveverilogdesignproblems}                       & 1  & 164   & Medium \\
m2 & poly\_interpolator~\cite{pinckney2025comprehensiveverilogdesignproblems}           & 2  & 224   & Medium \\
m3 & float\_multiplier~\cite{samidhm_float}                           & 2 & 218   & Medium \\
m4 & spi\_complex\_mult~\cite{pinckney2025comprehensiveverilogdesignproblems}           & 2  & 348   & Medium \\
m5 & rgb\_color\_space\_conv~\cite{pinckney2025comprehensiveverilogdesignproblems} & 1 & 282   & Medium \\
m6 & pooling~\cite{utlca_pooling}                                     & 1  & 412   & Medium \\
m7 & cryptech\_uart~\cite{secworks_uart}            & 2  & 447   & Medium \\
h1 & float\_adder~\cite{samidhm_float}                                & 2 & 463   & Hard \\
h2 & sha1\_top~\cite{secworks_sha1}                 & 3  & 630   & Hard \\
h3 & chacha\_top~\cite{secworks_chacha}             & 3  & 778  & Hard \\
h4 & trng\_csprng~\cite{secworks_trng}              & 3  & 1579  & Hard \\
h5 & trng\_mixer~\cite{secworks_trng}               & 3  & 2004  & Hard \\
h6 & trng\_top~\cite{secworks_trng}                 & 5 & 4099  & Hard \\
h7 & sd\_controller\_top~\cite{mczerski_sdcard}     & 3  & 4492  & Hard \\
\hline
\end{tabular}
\label{tab:design-difficulty}
\end{table}

\textbf{Evaluation Metrics:}
We employed multiple metrics to assess syntactic correctness of LLM's generated code,  adequacy of the generated testcase for coverage, and efficiency of running the framework:
\begin{enumerate}
    \item \textbf{pass@k:} Probability that at least one of $k$ generated testcases compiles and simulates correctly.
    \item \textbf{Coverage score:} Proportion of lines/statements exercised by a testcase. 
    \item \textbf{Token usage:} Number of tokens used per design.
    \item \textbf{Generation time:} Time spent on LLM generation.
\end{enumerate}

\textbf{Conversation Setup:}
To account for variability in LLM behavior, including occasional hallucinations or inconsistent outputs, for our evaluation, we conduct 5 independent conversations and compute the average coverage across them.
Up to 20 iterations are run per conversation (less than 20 if 100\% coverage is achieved). Coverage results are  merged across iterations within each conversation, reflecting the common practice in industry to merge coverage across all testcases for a design.
Also, the first testcase generated is simulated with 20 random seeds before moving on to iteratively closing coverage.
For experiments with generating a testplan enabled, the LLM generates the testplan before the coverage closure phase.


\section{Results} \label{sec:results}


\subsection{Overall Coverage Across Designs}
We first report the overall coverage achieved for each design under the mainline Spec2Cov configuration where testplan creation, batch generation, and context pruning are on.
To the best of our knowledge, there is no prior work that addresses automated testcase generation from specifications for closing code coverage for CVDP designs. Consequently, our results cannot be compared against a prior baseline, and we instead report absolute performance metrics for our framework.

\begin{table}[t]
\centering
\caption{Spec2Cov achieves high coverage across benchmarks, with a 91.47\% geometric mean. Coverage values represent the best performance achieved by merging results across multiple independent runs.}
\begin{tabular}{lc|lc}
\textbf{Design} & \textbf{Spec2Cov} & \textbf{Design} & \textbf{Spec2Cov} \\
\hline
lfsr & 100 & poly\_interpolator & 100 \\
caesar\_cipher & 80.00 & rgb\_conversion & 100 \\
dual\_port\_memory & 100 & spi\_complex\_mult & 100 \\
multiplexer & 100 & pooling & 97.71\\
cont\_adder & 100 & cryptech\_uart* & 100 \\
alu & 100 & float\_adder & 91.58 \\
fixed\_arbiter & 100 & sha1\_top & 100 \\
vndecorrelator & 100 & chacha\_top & 96.54 \\
fifo & 100 & trng\_csprng & 82.67 \\
ttc\_lite & 83.33 & trng\_mixer & 76.78 \\
memory\_scheduler & 98.28 & trng\_top & 55.55 \\
door\_lock & 100 & sd\_controller\_top & 49.31 \\
sorter & 100 & \textbf{geomean} & 91.47 \\
float\_multiplier & 100 &  &  \\
 \hline
\end{tabular}
\label{tab:mainline_results}
    \vspace{-3mm}
\end{table}

The results are shown in Table \ref{tab:mainline_results} 
with designs ordered by size. Low-difficulty designs (simple datapath or control structures) achieve near-100\% coverage on average. Medium-difficulty designs (moderate complexity, such as stateful controllers or multi-module arithmetic units) average between 85--95\% coverage. High-difficulty designs (cryptographic cores and large streaming datapaths) remain below 80--90\% coverage on average, with the lowest observed coverage at 49\%. This trend highlights both the strength of LLMs on simpler designs and the remaining verification challenges in scaling toward industrial-scale designs.

\subsection{Effect of Agentic Approach}
\begin{figure}[t]
    \centering
    \includegraphics[width=\linewidth]{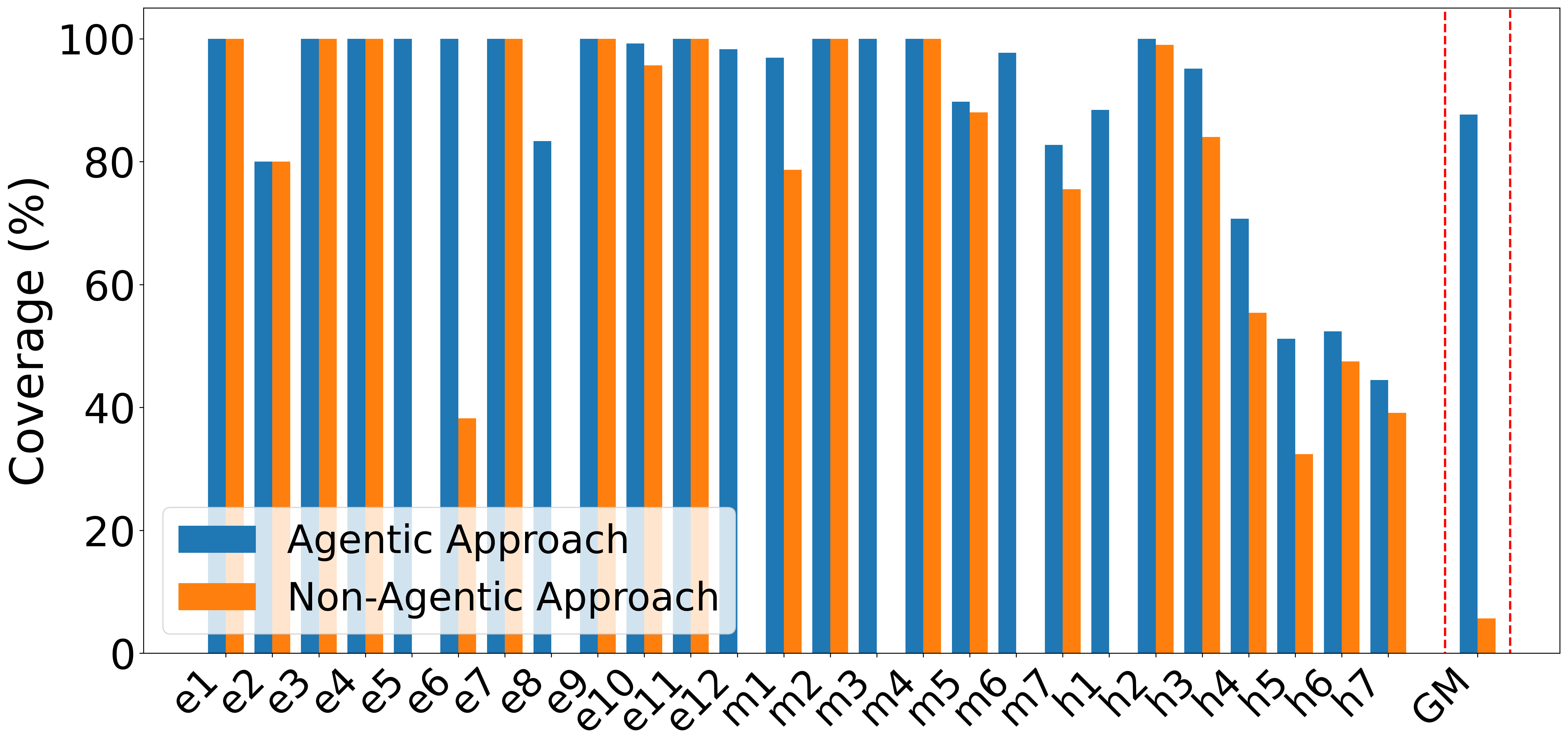}
    \caption{Agentic approach achieves significantly higher coverage, whereas the single iteration approach fails to produce any coverage in some cases. GM = Geometric Mean}
    \label{fig:single-shot-v-agentic}
    \vspace{-2mm}
\end{figure}

We illustrate the benefit of iterative prompting by comparing a single testcase generation to an agentic workflow with 20 iterations. 
Fig. \ref{fig:single-shot-v-agentic} shows that iterative, feedback-driven prompting substantially improves coverage over one-shot generation.

\subsection{Effect of Batched Generation, Testplan Creation, and Context Pruning} \label{sec:ablation}

\begin{figure}
    \centering
    \begin{subfigure}{\linewidth}
        \centering
        \includegraphics[width=\textwidth]{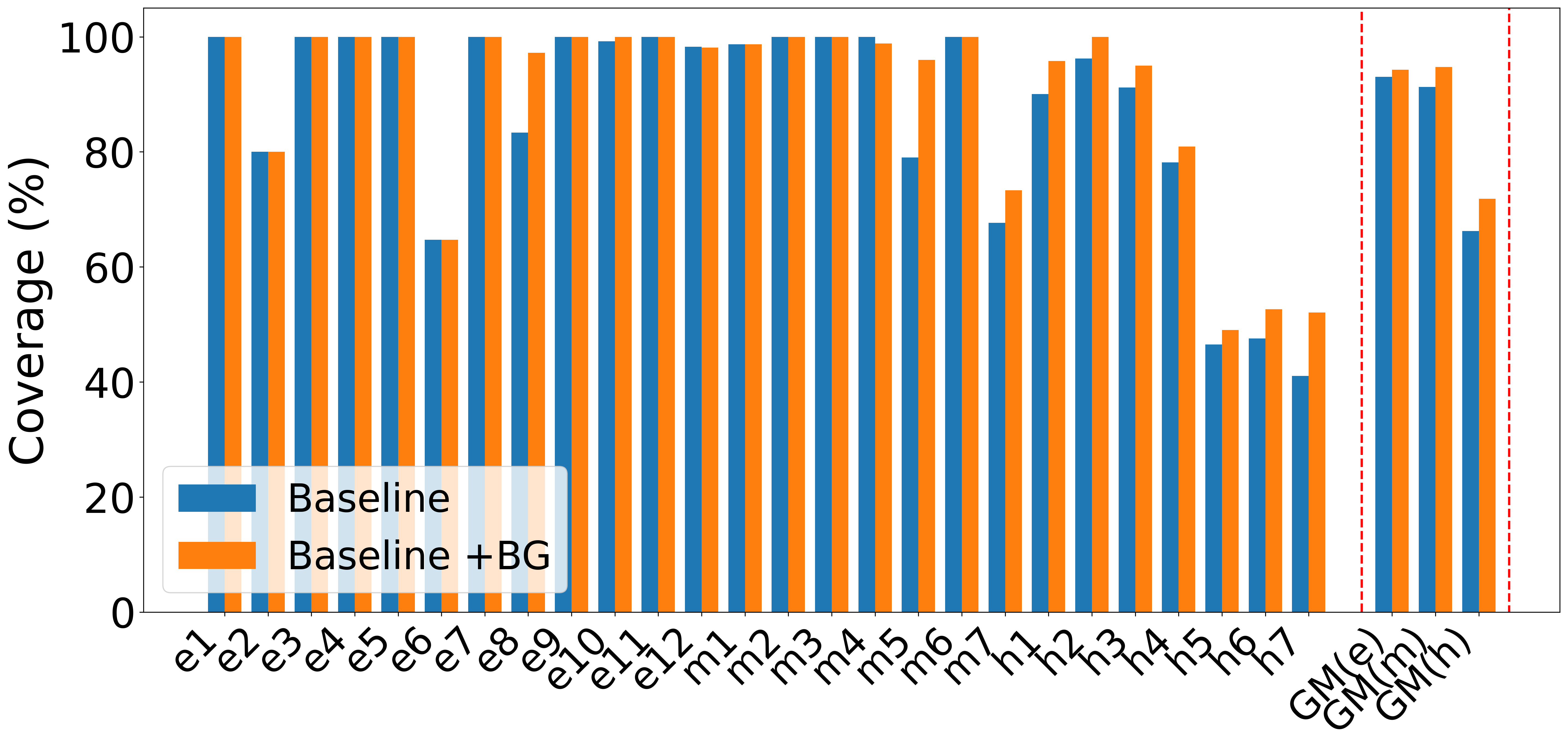}
    \end{subfigure}    
    \begin{subfigure}{\linewidth}
        \centering
        \includegraphics[width=\textwidth]{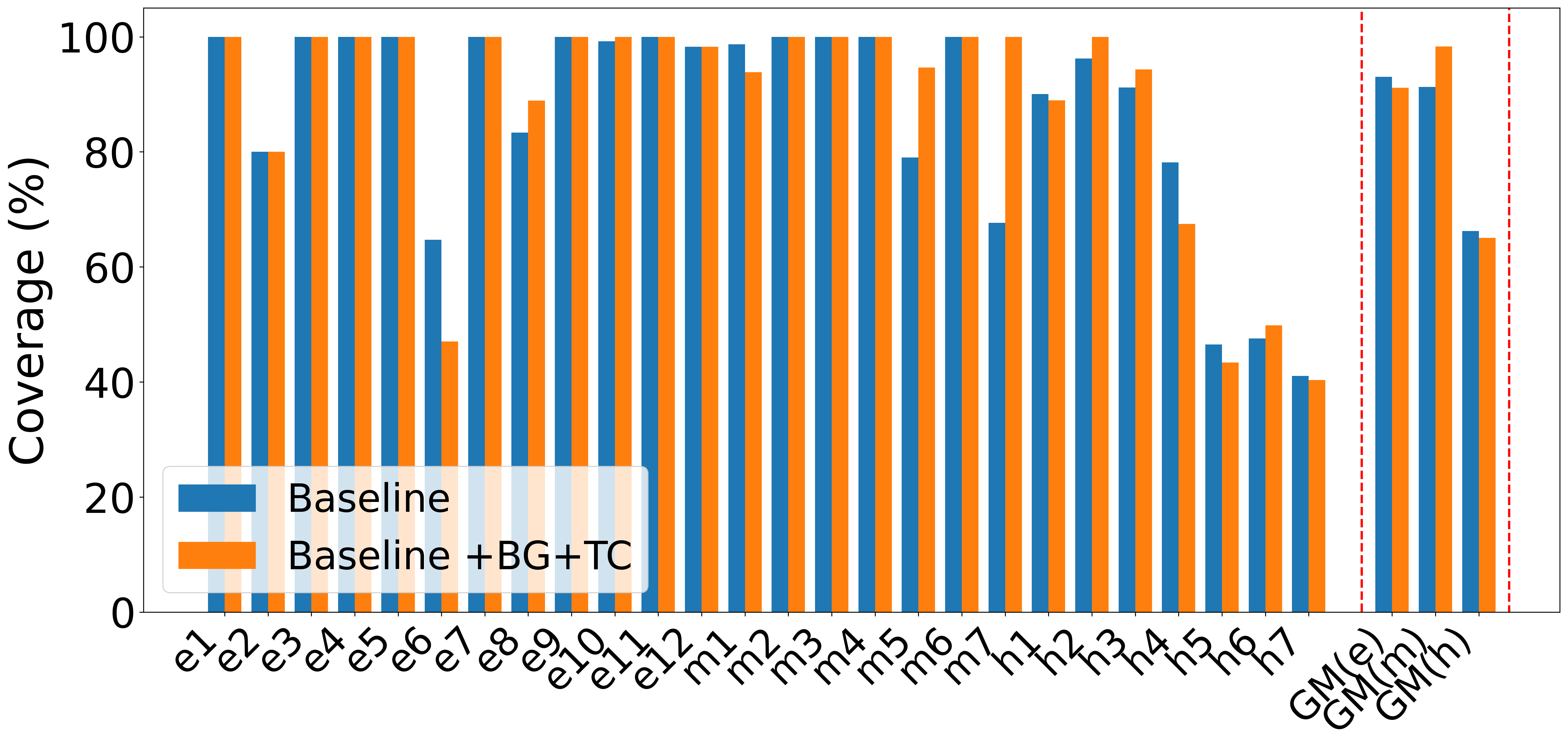}
    \end{subfigure}    
    \begin{subfigure}{\linewidth}
        \centering
        \includegraphics[width=\textwidth]{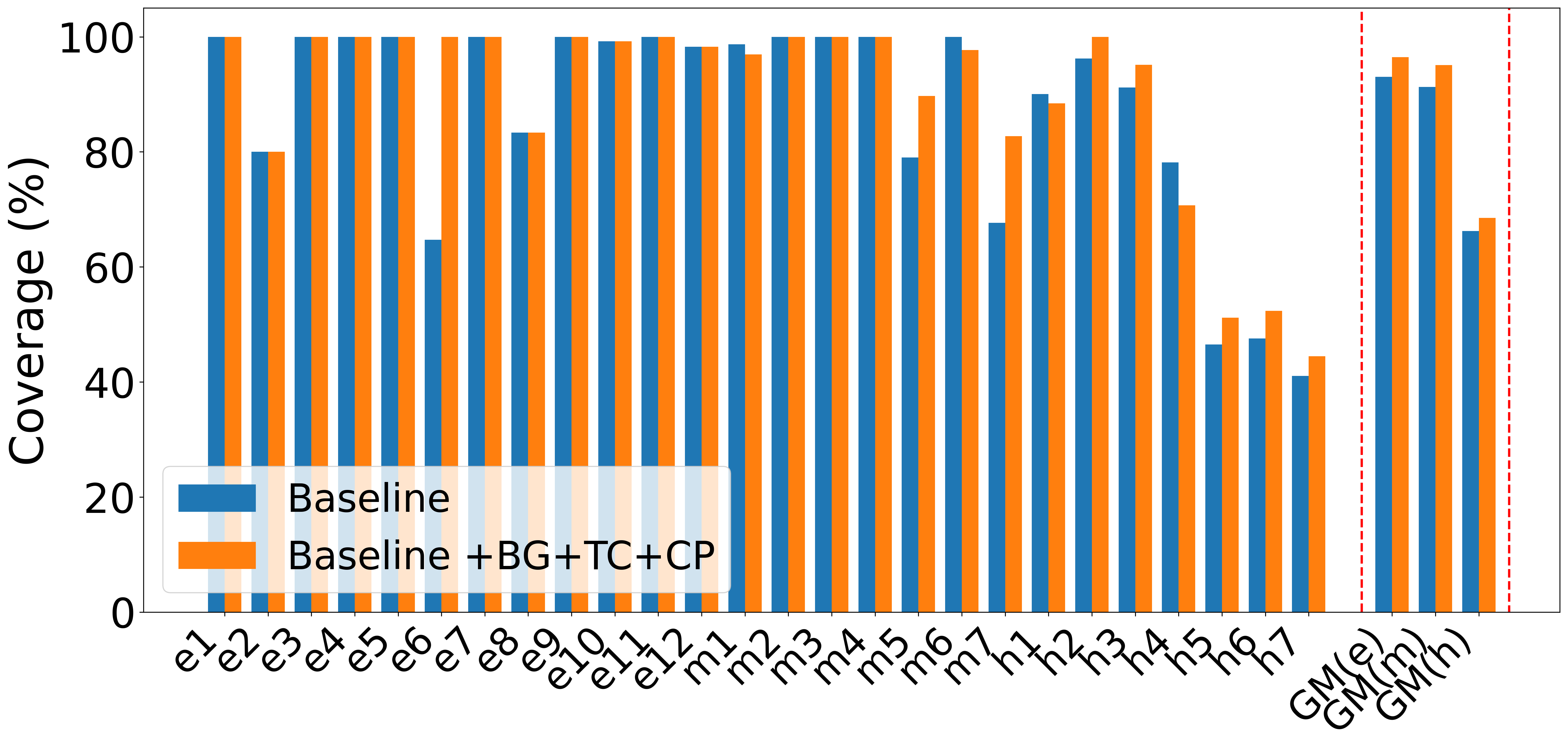}
    \end{subfigure}    
    \caption{Impact on code coverage achieved by progressively adding Batched Generation (BG), Testplan Creation (TC), and Context Pruning (CP) features. All figures compare against the baseline. GM(x) = Geometric Mean of e(easy), m(medium), h(hard) designs.}
    \label{fig:ablation}
\end{figure}

We conduct studies to isolate the effect of key features: batched generation (BG), testplan creation (TC), and context pruning (CP).
We start with a baseline that does not have any of these features enabled, and then progressively add them in the order BG, TC, and CP to observe their cumulative impact.
The results are presented in Fig. \ref{fig:ablation}, wherein geometric means are presented across easy designs (GM(e)), medium designs (GM(m)), and hard designs (GM(h)).
Unlike Table \ref{tab:mainline_results}, which reports the best coverage achieved by merging across runs, these figures show the average coverage achieved by a single conversation to illustrate typical single-run performance.




The top subfigure shows that batched generation  modestly improves medium and hard designs, while easy designs already achieve high coverage without this feature.

The middle subfigure reveals that adding testplan creation on top of batched generation does not consistently improve coverage. For some hard designs, this combination actually degrades performance compared to the baseline, suggesting that the testplan feature requires further refinement.
To investigate this, we devise an enhanced testplan approach in which the LLM generates a feature-based testplan, then creates independent testcases targeting each feature before moving to iterative coverage closure (Phase 2). Preliminary results on selected designs are shown in Table \ref{tab:feature_based_results}, demonstrating promise for hierarchical feature-targeted testing that will guide future enhancements.

\begin{table}[t]
\centering
\caption{Enhanced feature-based testplan approach leads to significant improvement over the baseline.}
\begin{tabular}{lccc}
\textbf{Design} & \textbf{Baseline} & \textbf{Enhanced Testplan} & \textbf{$\Delta$} \\
\hline
caesar\_cipher~\cite{pinckney2025comprehensiveverilogdesignproblems} & 80.00 & 80.00 & +0.00 \\
fixed\_arbiter~\cite{pinckney2025comprehensiveverilogdesignproblems} & 100 & 100.00 & +0.00 \\
ttc\_lite~\cite{pinckney2025comprehensiveverilogdesignproblems} & 83.33 & 100.00 & +16.67 \\
pooling~\cite{utlca_pooling} & 97.71 & 99.54 & +1.83 \\
trng\_top~\cite{secworks_trng} & 55.55 & 70.88 & +15.38 \\
\hline
\end{tabular}
\label{tab:feature_based_results}
\vspace{-2mm}
\end{table}

The bottom subfigure shows the complete mainline configuration with all three features enabled (BG+TC+CP). Notably, context pruning provides substantial recovery from the degradation caused by testplan creation, with some designs (e.g., fixed\_arbiter) achieving dramatic improvements to 100\% coverage. This demonstrates that while individual features may not always help in isolation, their combination in the mainline configuration delivers strong overall performance. Context pruning is particularly crucial, as it helps the LLM focus on relevant information by removing polluted context, which can improve testcase generation quality.

Spec2Cov provides extensive configurability 
to enable/disable features from the command line. Users can try multiple configurations for new designs to determine what works best.


\subsection{Compilation and Simulation Success}
We report pass@1/3/5 across various designs in Table \ref{tab:passk}, capturing compilation and simulation success. We also show the effect of removing features on pass@k metrics.
Removing testplan creation gives the highest pass@1, which was not an expected outcome. It is possible that the LLM tries to satisfy the entire testplan in the first iteration, leading to one or more errors, while in other configurations the LLM starts slower. 
Future work will investigate this behavior using a white-box approach. Removing the batch generation feature gives the highest pass@3 and pass@5, which makes intuitive sense because this feature enables generating more variations that fail, but also more variations that obtain higher coverage. Because of long-context degradation experienced by LLMs, where an LLM performs worse with too long of a context length, removing context pruning results in lower pass@k scores compared to other features.

\begin{table}[t]
\centering
\caption{Spec2Cov achieves competitive pass@1 while maintaining stable performance at pass@3 and pass@5, demonstrating its robustness compared to feature-ablated configurations.}
\label{tab:passk}
\begin{tabular}{lccc}
\hline
\textbf{Configuration} & \textbf{pass@1} & \textbf{pass@3} & \textbf{pass@5} \\
\hline
Spec2Cov   & 0.500 & 0.565 & 0.609 \\
--Testplan Creation & \textbf{0.587} & 0.623 & 0.638 \\
--Batched Generation    & 0.420 & \textbf{0.870} & \textbf{0.891} \\
--Context Pruning      & 0.354 & 0.507 & 0.632 \\
\hline
\end{tabular}
\vspace{-3mm}
\end{table}

The CVDP paper \cite{pinckney2025comprehensiveverilogdesignproblems}
presents pass@1 data across eight LLMs. However, since the LLM used we use (ChatGPT-4o) is not included in the CVDP paper, we cannot make a direct comparison. However, to perform an indirect comparison, we can average the pass@1 across OpenAI models (ChatGPT-4.1, ChatGPT-o1, and ChatGPT-o1-mini) for problem sets cid12-14, which is 0.112. Thus, there is an improvement on CVDP verification problems using Spec2Cov than when using CVDP's default agentic method.

\subsection{Generation Costs and Efficiency}
Finally, we evaluate the cost of using the LLM in Spec2Cov using generation time and token count. The geometric means for these metrics across hard, medium and easy design are shown in Fig.~\ref{fig:cost}. 
Although token count and generation time generally increase with design size, anomalies occur because they also depend on iteration count and coverage difficulty.
Table \ref{tab:num_iters} shows the number of iterations taken by each design to achieve the coverage shown in Table \ref{tab:mainline_results}.
Each iteration corresponds to a single LLM-simulator exchange.
Because conversations run in parallel as independent processes, Spec2Cov can evaluate many design-testbench interactions quickly, reducing wall-clock time for coverage exploration.
While traditional verification workflows often require hours or  days to hand-craft and refine testcases, our agentic approach completes comparable exploration in minutes, demonstrating scalability for large-scale hardware verification.

\begin{table}[h]
\begin{minipage}{0.22\textwidth}
    \centering
    \begin{tabular}{p{0.5cm}p{0.5cm}|p{0.5cm}p{0.5cm}}
    \textbf{Design} & \textbf{\#Iter.} & \textbf{Design} & \textbf{\#Iter.} \\
    \hline
    e1 & 2 & m1 & 20 \\
    e2 & 20 & m2 & 1 \\
    e3 & 2 & m3 & 2 \\
    e4 & 1 & m4 & 1 \\
    e5 & 2 & m5 & 20 \\
    e6 & 11 & m6 & 20\\
    e7 & 1 & m7 & 20 \\
    e8 & 20 & h1 & 20 \\
    e9 & 1 & h2 & 20 \\
    e10 & 20 & h3 & 20 \\
    e11 & 1 & h4 & 20 \\
    e12 & 20 & h5 & 20 \\
     &  & h6 & 20 \\
     &  & h7 & 20 \\
     \hline
    \end{tabular}
    \caption{Average number of iterations required for each design to achieve coverage shown in Table \ref{tab:mainline_results}.}
    \label{tab:num_iters}
\end{minipage}\hfill
\begin{minipage}{0.25\textwidth}
    \centering
    \includegraphics[width=\linewidth]{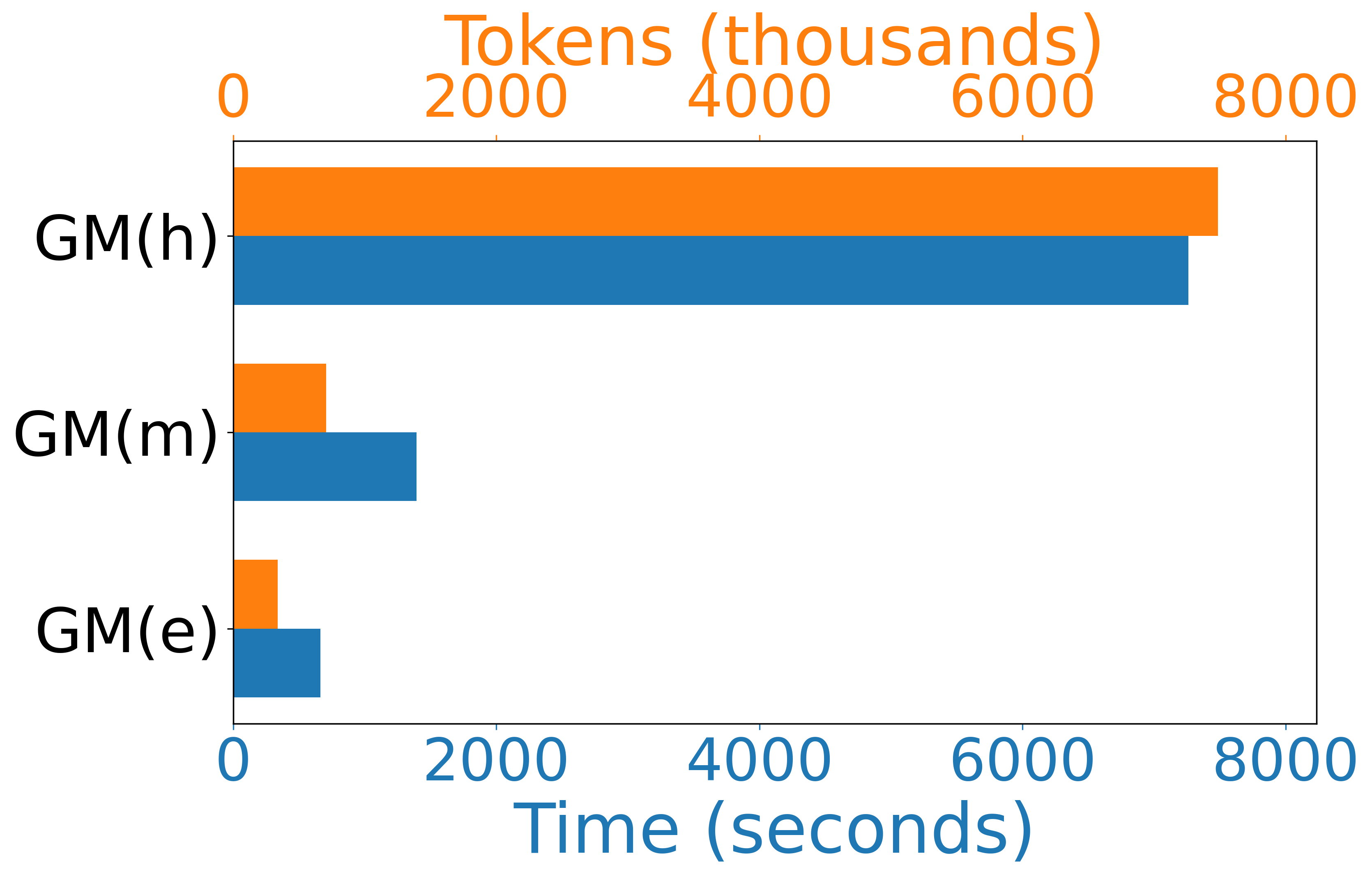}
    \captionof{figure}{Average generation time and average total token usage increases as design complexity increases. }
    \label{fig:cost}
\end{minipage}
    \vspace{-5mm}
\end{table}

\section{Discussion} \label{sec:discussion}



\textbf{Code Coverage Focus}
Spec2Cov targets code coverage because it is a standard sign-off gate before broader verification closure, and manual closure remains a major verification bottleneck. Automating this stage provides immediate workflow impact and establishes a foundation for future functional-coverage extensions.

\textbf{LLM-Based Stimulus Generation}
Spec2Cov uses LLMs because scripted and constrained-random methods alone often miss corner-case behavior without manual guidance. LLMs add context-aware testplan generation, reasoning-based testcase development and iterative coverage enhancement, reducing engineer effort in coverage closure.

\textbf{Scalability}
Spec2Cov is designed for module- and subsystem-level verification in hierarchical flows, and our evaluation includes designs up to five hierarchy levels and 4500 lines of code. Future work will include RAG-based context management, hierarchical test composition, and feature-based testplanning (early evidence in Section \ref{sec:ablation}).

\textbf{Rich-Text Specifications}
Most specification documents are written in rich-text format with tables and figures. Spec2Cov supports ingesting feature-rich PDFs and extracts content by using PyMuPDF and pdfplumber tools.

\section{Conclusion} \label{sec:conclusion}


We present Spec2Cov, an agentic framework that automates code coverage closure for digital hardware designs by generating testcases directly from specifications and iteratively refining them using simulator feedback. 
Our evaluation demonstrates that an LLM-driven, tool-integrated workflow can substantially reduce manual effort in verification while maintaining competitive coverage and compilation success rates. 

\bibliographystyle{IEEEtran}
\bibliography{refs}

@IEEEtranBSTCTL{BSTcontrol,
  CTLdash_repeated_names = "no",
}

@inproceedings{blocklove:chipchat:2023,
  title = {Chip-{{Chat}}: {{Challenges}} and {{Opportunities}} in {{Conversational Hardware Design}}},
  shorttitle = {Chip-{{Chat}}},
  booktitle = {2023 {{ACM}}/{{IEEE}} 5th {{Workshop}} on {{Machine Learning}} for {{CAD}} ({{MLCAD}})},
  author = {Blocklove, Jason and Garg, Siddharth and Karri, Ramesh and Pearce, Hammond},
  year = {2023},
  month = sep,
  eprint = {2305.13243},
  primaryclass = {cs},
  pages = {1--6},
  doi = {10.1109/MLCAD58807.2023.10299874},
  urldate = {2023-12-20},
  archiveprefix = {arXiv}
}

@article{thakur:verigen:2024,
  title = {{{VeriGen}}: {{A Large Language Model}} for {{Verilog Code Generation}}},
  shorttitle = {{{VeriGen}}},
  author = {Thakur, Shailja and Ahmad, Baleegh and Pearce, Hammond and Tan, Benjamin and {Dolan-Gavitt}, Brendan and Karri, Ramesh and Garg, Siddharth},
  year = {2024},
  month = feb,
  journal = {ACM Transactions on Design Automation of Electronic Systems},
  pages = {3643681},
  issn = {1084-4309, 1557-7309},
  doi = {10.1145/3643681},
  urldate = {2024-03-23},
  langid = {english}
}

@inproceedings{mali:chiraag:2024a,
  title = {{{ChIRAAG}}: {{ChatGPT Informed Rapid}} and {{Automated Assertion Generation}}},
  shorttitle = {{{ChIRAAG}}},
  booktitle = {2024 {{IEEE Computer Society Annual Symposium}} on {{VLSI}} ({{ISVLSI}})},
  author = {Mali, Bhabesh and Maddala, Karthik and Gupta, Vatsal and Reddy, Sweeya and Karfa, Chandan and Karri, Ramesh},
  year = {2024},
  month = jul,
  pages = {680--683},
  publisher = {IEEE},
  address = {Knoxville, TN, USA},
  doi = {10.1109/ISVLSI61997.2024.00130},
  urldate = {2024-11-04},
  isbn = {9798350354119},
  langid = {english}
}

@inproceedings{fang:assertllm:2024a,
  title = {{{AssertLLM}}: {{Generating Hardware Verification Assertions}} from {{Design Specifications}} via {{Multi-LLMs}}},
  shorttitle = {{{AssertLLM}}},
  booktitle = {2024 {{IEEE LLM Aided Design Workshop}} ({{LAD}})},
  author = {Fang, Wenji and Li, Mengming and Li, Min and Yan, Zhiyuan and Liu, Shang and Zhang, Hongce and Xie, Zhiyao},
  year = {2024},
  month = jun,
  pages = {1--1},
  doi = {10.1109/LAD62341.2024.10691792},
  urldate = {2024-11-04}
}

@inproceedings{ma:verilogreader:2024a,
  title = {{{VerilogReader}}: {{LLM-Aided Hardware Test Generation}}},
  shorttitle = {{{VerilogReader}}},
  booktitle = {2024 {{IEEE LLM Aided Design Workshop}} ({{LAD}})},
  author = {Ma, Ruiyang and Yang, Yuxin and Liu, Ziqian and Zhang, Jiaxi and Li, Min and Huang, Junhua and Luo, Guojie},
  year = {2024},
  month = jun,
  pages = {1--5},
  doi = {10.1109/LAD62341.2024.10691801},
  urldate = {2024-11-04}
}

@misc{zhang:llm4dv:2023,
  title = {{{LLM4DV}}: {{Using Large Language Models}} for {{Hardware Test Stimuli Generation}}},
  shorttitle = {{{LLM4DV}}},
  author = {Zhang, Zixi and Chadwick, Greg and McNally, Hugo and Zhao, Yiren and Mullins, Robert},
  year = {2023},
  month = oct,
  number = {arXiv:2310.04535},
  eprint = {2310.04535},
  primaryclass = {cs},
  publisher = {arXiv},
  doi = {10.48550/arXiv.2310.04535},
  urldate = {2024-06-28},
  archiveprefix = {arXiv}
}

@inproceedings{qiu:autobench:2024a,
  title = {{{AutoBench}}: {{Automatic Testbench Generation}} and {{Evaluation Using LLMs}} for {{HDL Design}}},
  shorttitle = {{{AutoBench}}},
  booktitle = {Proceedings of the 2024 {{ACM}}/{{IEEE International Symposium}} on {{Machine Learning}} for {{CAD}}},
  author = {Qiu, Ruidi and Zhang, Grace Li and Drechsler, Rolf and Schlichtmann, Ulf and Li, Bing},
  year = {2024},
  month = sep,
  series = {{{MLCAD}} '24},
  pages = {1--10},
  publisher = {Association for Computing Machinery},
  address = {New York, NY, USA},
  doi = {10.1145/3670474.3685956},
  urldate = {2024-11-03},
  isbn = {9798400706998}
}

@misc{zhang2025llm4dvusinglargelanguage,
      title={ILLM4DV: Using Large Language Models for Hardware Test Stimuli Generation}}

@misc{ye2025conceptpracticeautomatedllmaided,
      title={{From Concept to Practice: an Automated LLM-aided UVM Machine for RTL Verification}}, 
      author={Junhao Ye and Yuchen Hu and Ke Xu and Dingrong Pan and Qichun Chen and Jie Zhou and Shuai Zhao and Xinwei Fang and Xi Wang and Nan Guan and Zhe Jiang},
      year={2025},
      eprint={2504.19959},
      archivePrefix={arXiv},
      primaryClass={cs.AR},
      url={https://arxiv.org/abs/2504.19959}, 
}

@misc{pinckney2025comprehensiveverilogdesignproblems,
      title={{Comprehensive Verilog Design Problems: A Next-Generation Benchmark Dataset for Evaluating Large Language Models and Agents on RTL Design and Verification}}, 
      author={Nathaniel Pinckney and Chenhui Deng and Chia-Tung Ho and Yun-Da Tsai and Mingjie Liu and Wenfei Zhou and Brucek Khailany and Haoxing Ren},
      year={2025},
      eprint={2506.14074},
      archivePrefix={arXiv},
      primaryClass={cs.LG},
      url={https://arxiv.org/abs/2506.14074}, 
}

@misc{fu2025gpt4aigchipnextgenerationaiaccelerator,
      title={{GPT4AIGChip: Towards Next-Generation AI Accelerator Design Automation via Large Language Models}}, 
      author={Yonggan Fu and Yongan Zhang and Zhongzhi Yu and Sixu Li and Zhifan Ye and Chaojian Li and Cheng Wan and Yingyan Celine Lin},
      year={2025},
      eprint={2309.10730},
      archivePrefix={arXiv},
      primaryClass={cs.LG},
      url={https://arxiv.org/abs/2309.10730}, 
}

@INPROCEEDINGS{hassan:promptverifyrepeat:2025,
  author={Hassan, Muhammad and Nadeem, Mohamed and Qayyum, Khushboo and Jha, Chandan Kumar and Drechsler, Rolf},
  booktitle={2025 IEEE International Conference on Omni-layer Intelligent Systems (COINS)}, 
  title={{Prompt. Verify. Repeat. LLMs in the Hardware Verification Cycle}}, 
  year={2025},
  volume={},
  number={},
  pages={1-6},
  doi={10.1109/COINS65080.2025.11125767}}

@misc{calzada2025verilogdblargesthighestqualitydataset,
      title={{VerilogDB: The Largest, Highest-Quality Dataset with a Preprocessing Framework for LLM-based RTL Generation}}, 
      author={Paul E. Calzada and Zahin Ibnat and Tanvir Rahman and Kamal Kandula and Danyu Lu and Sujan Kumar Saha and Farimah Farahmandi and Mark Tehranipoor},
      year={2025},
      eprint={2507.13369},
      archivePrefix={arXiv},
      primaryClass={cs.AR},
      url={https://arxiv.org/abs/2507.13369}, 
}

@inproceedings{kwon2023efficient,
  title={{Efficient Memory Management for Large Language Model Serving with PagedAttention}},
  author={Woosuk Kwon and Zhuohan Li and Siyuan Zhuang and Ying Sheng and Lianmin Zheng and Cody Hao Yu and Joseph E. Gonzalez and Hao Zhang and Ion Stoica},
  booktitle={Proceedings of the ACM SIGOPS 29th Symposium on Operating Systems Principles},
  year={2023}
}

@inbook{xu:meic:2025,
author = {Xu, Ke and Sun, Jialin and Hu, Yuchen and Fang, Xinwei and Shan, Weiwei and Wang, Xi and Jiang, Zhe},
title = {{MEIC: Re-thinking RTL Debug Automation using LLMs}},
year = {2025},
isbn = {9798400710773},
publisher = {Association for Computing Machinery},
address = {New York, NY, USA},
url = {https://doi-org.ezproxy1.lib.asu.edu/10.1145/3676536.3676801},
booktitle = {Proceedings of the 43rd IEEE/ACM International Conference on Computer-Aided Design},
articleno = {100},
numpages = {9}
}

@misc{hu2024uvllmautomateduniversalrtl,
      title={{UVLLM: An Automated Universal RTL Verification Framework using LLMs}}, 
      author={Yuchen Hu and Junhao Ye and Ke Xu and Jialin Sun and Shiyue Zhang and Xinyao Jiao and Dingrong Pan and Jie Zhou and Ning Wang and Weiwei Shan and Xinwei Fang and Xi Wang and Nan Guan and Zhe Jiang},
      year={2024},
      eprint={2411.16238},
      archivePrefix={arXiv},
      primaryClass={cs.AR},
      url={https://arxiv.org/abs/2411.16238}, 
}

@misc{spec2cov_anonymized,
author = "Anonymous",
title = {{Spec2Cov}},
note = {\url{https://anonymous.4open.science/r/spec2cov}},
year={2025}
}

@misc{secworks_vndecorrelator,
  author = {Joachim Str{\"o}mbergson},
  title = {{vndecorrelator: Verilog implementation of a von Neumann decorrelator}},
  howpublished = {\url{https://github.com/secworks/vndecorrelator}},
  year = {2016}
}

@misc{avashist_fifo,
  author = {Ansh Vashist},
  title = {{FIFO\_SystemVerilog\_Assertion: Synchronous FIFO with SystemVerilog Assertions}},
  howpublished = {\url{https://github.com/avashist003/FIFO\_SystemVerilog\_Assertion}},
  year = {2020}
}

@misc{secworks_uart,
  author = {Joachim Str{\"o}mbergson},
  title = {{uart: Verilog implementation of a simple UART core}},
  howpublished = {\url{https://github.com/secworks/uart}},
  year = {2014}
}

@misc{secworks_sha1,
  author = {Joachim Str{\"o}mbergson},
  title = {{sha1: Verilog implementation of the SHA-1 hash function}},
  howpublished = {\url{https://github.com/secworks/sha1}},
  year = {2014}
}

@misc{secworks_chacha,
  author = {Joachim Str{\"o}mbergson},
  title = {{chacha: Verilog implementation of the ChaCha stream cipher}},
  howpublished = {\url{https://github.com/secworks/chacha}},
  year = {2014}
}

@misc{secworks_trng,
  author = {Joachim Str{\"o}mbergson},
  title = {{trng: True Random Number Generator core implemented in Verilog}},
  howpublished = {\url{https://github.com/secworks/trng}},
  year = {2014}
}

@misc{mczerski_sdcard,
  author = {Marek Czerski},
  title = {{SD-card-controller: SD/SDHC card controller for Wishbone bus}},
  howpublished = {\url{https://github.com/mczerski/SD-card-controller}},
  year = {2013}
}

@misc{samidhm_float,
  author = {Samidh Mehta},
  title = {{DSP\_Slice: Floating Point Units}},
  howpublished = {\url{https://github.com/samidhm/DSP\_Slice/tree/main/Floating\_Point\_Units}},
  year = {2020}
}

@misc{utlca_pooling,
  author = {Vedant Patel and UT-LCA},
  title = {{tpu\_like\_design: TPU-like design with pooling unit}},
  howpublished = {\url{https://github.com/UT-LCA/tpu\_like\_design/tree/master/design\_ws\_vedant}},
  year = {2019}
}

\end{document}